\documentclass[twocolumn,pra,unsortedaddress,showpacs,floatfix,
citeautoscript,nofootinbib]{revtex4-2}
\usepackage{amsmath,amssymb,mathrsfs}

\usepackage{amsbsy}
\usepackage{psfrag}
\usepackage{graphicx}
\usepackage{epsfig}
\usepackage{bm}
\usepackage{color}

\usepackage[normalem]{ulem}
\usepackage{latexsym}
\usepackage[colorlinks,urlcolor=blue,citecolor=blue]{hyperref}
\usepackage{dcolumn}
\usepackage{subfigure}
\def\sumint{\int \!\!\!\!\!\!\!\!\! \sum }
\def\sint{\int \!\!\!\!\!\! \sum }

\begin{document}

\title [Low-Dimensional Self-Bound Quantum Rabi-Coupled Bosonic Droplets]
{Low-Dimensional Self-Bound Quantum Rabi-Coupled Bosonic Droplets}

\author{Emerson Chiquillo}
\affiliation{Escuela de F\'isica, Universidad Pedag\'ogica y Tecnol\'ogica de 
Colombia (UPTC),\\
Avenida Central del Norte, 150003 Tunja, Colombia}
% \ead{emerson.chiquillo@uptc.edu.co}

\begin{abstract}
We analytically calculate the leading quantum corrections of the ground-state
energy of two- and one-dimensional weakly interacting Rabi-coupled Bose-Bose 
mixtures in the frame of the Bogoliubov approximation.
We show that to repulsive intraspecies and attractive interspecies interactions,
the effect of quantum fluctuations favors the formation of self-bound droplets. 
These liquidlike states are crucially affected by the Rabi coupling, leading thus
to the appearance of a quantum instability. 
We derive meaningful formulas to describe the droplet phase in the 
one-dimensional case. 
\end{abstract}

\maketitle
\textit{Introduction.}
In the last few years, self-bound states in ultracold and ultradilute gases
beyond the mean-field approximation have attracted the attention providing
new opportunities for experimental and theoretical research.
In the pioneering theoretical works \cite{Petrov1,Petrov2}, 
the quantum fluctuations have been proposed as the main stabilization mechanism
of such self-bound structures in Bose-Bose mixtures, revealing the crucial role
played by quantum many-body effects.
This stabilization lies in the balance of the mean-field contribution, close to
the collapse threshold, and the first quantum correction coming from 
the zero-point motion of the Bogoliubov excitations, allowing thus the birth of 
a new phase of matter, called liquidlike quantum droplet.
Soon after this prediction, and in a different context, droplets in
dipolar gases of atoms of $^{164}$Dy 
\cite{Rosensweig,Self-Drop-Dy,Trap-Drop-Dy-1,Trap-Drop-Dy-2} and $^{168}$Er 
\cite{Erbium} were discovered.
These intriguing breakthroughs reveal that strongly bosonic gases of magnetic
atoms do not necessarily collapse, as previously assumed from a purely mean-field
viewpoint. Instead, they get into a dipolar droplet phase.
More recently, the predictions in Ref. \cite{Petrov1} have been experimentally 
achieved with ultracold and dilute mixtures of gases of $^{39}$K atoms 
\cite{Exper-droplet1,Exper-droplet2,Exper-droplet3}. 
From a theoretical viewpoint, such a droplet phase remains ultradilute and 
weakly interacting, allowing for a perturbative description \cite{LHY-1,LHY-2}.
In order to explain the experimental results obtained with dipolar Bose gases, 
a generalized time-dependent nonlocal Gross-Pitaevskii equation has been solved
\cite{Three-body-1,Three-body-2,trapped-dip-LHY-1,trapped-dip-LHY-2,d-LHY-1,
d-LHY-2}. 
There have also been numerical works 
\cite{Montecarlo-dip1,Montecarlo-dip2,Montecarlo-dip3}.
In the study of the leading quantum corrections in Bose-Bose mixtures, works 
employing Monte Carlo methods have also been carried out 
\cite{Montecarlo-Bose-Bose1,Montecarlo-Bose-Bose2}.
Thermal corrections on mixtures of two-component bosonic gases with short-range
interactions have been studied in Ref. \cite{1D-3D-mixtures}.

On the other hand, in the last decade, laser beams have been used to induce 
artificial transitions among different atomic hyperfine states \cite{S1,S2}.
In the framework of the mean-field approximation, extensive research was 
addressed to understand the properties of these synthetic non-Abelian gauge
fields in neutral bosonic mixtures of ultracold gases \cite{S3,S4,S5}.
These works have opened the door to a fascinating and fast development of
phenomena with spin-orbit- and Rabi-coupled ultracold atoms \cite{S6}.
Recently, a connection between the synthetic Rabi coupling and the quantum
droplets in a three-dimensional (3D) two-component Bose gas of interacting
alkali-metal atoms was investigated in Ref. \cite{Rabi}.
However, the liquidlike droplet phase is more ubiquitous and remarkable in 
low-dimensional Bose-Bose mixtures, as predicted in Ref. \cite{Petrov2}.
Thus these systems have given rise to great experimental and theoretical interest.
Then, for the best understanding of the behavior of interacting pseudospinor
bosonic systems in a lower dimensionality, it is important to analyze the quantum 
effect induced by the Rabi coupling.

In this Rapid Communication, motivated by the enhanced role of beyond-mean-field 
effects in ultracold gases, and the fast development of artificial couplings 
between atomic internal states, we address theoretically the formation and
stability of self-bound liquidlike droplets in low-dimensional Rabi-coupled 
ultracold bosonic atoms.
We consider two- and one-dimensional two-component mixtures, and we focus on the 
interesting case where the intraspecies interactions are weakly repulsive and 
the interspecies ones are weakly attractive.
We obtain the conditions of formation of self-bound droplets in terms of the Rabi
coupling and the strength of the interactions.
Remarkably, we find that in both two- and one-dimensional scenarios, there is a
critical Rabi frequency beyond which the self-bound droplet becomes unstable.
In the one-dimensional (1D) case, and by considering a small Rabi-coupling 
regime, we obtain meaningful analytical formulas to describe the ground state of
the droplet phase.
It is also relevant to stress that such a 1D instability is similar to the 
collapse in a 3D Bose-Einstein condensate (BEC).
At equilibrium some relevant quantities are also calculated, such as the chemical
potential and speed of sound.

\textit{Rabi-coupled bosons.}
Consider the path-integral formalism for two interacting and equal-mass 
Rabi-coupled bosonic species with hyperfine states $(\uparrow,\downarrow)$, and 
governed by the action
\begin{eqnarray}
S[\Psi,\Psi^*]&=&\sumint \Big[\psi_\alpha ^*\Big(\hbar\frac{\partial}
{\partial\tau} - \frac{\hbar^2}{2m} \nabla^2  - \mu \Big)\psi_\alpha 
\nonumber \\
&+& \frac{1}{2}\sum_{\sigma}g_{\alpha\sigma}
|\psi_\alpha|^2|\psi_\sigma|^2 - \hbar\omega_R(\psi_\uparrow^*\psi_\downarrow 
- \psi_\downarrow^*\psi_\uparrow)\Big],
\end{eqnarray}
where we have used the shorthand notation 
$\,\sint \equiv\int_0^{\hbar \beta} d\tau \int_{L^D}  d^D r
\sum_{\alpha}$, $\beta = 1/k_B T$, $k_B$ is the Boltzmann constant, $D=1,2$, 
and $\alpha,\sigma=\uparrow,\downarrow$.
Given the pseudospinor $\Psi=(\psi_\uparrow,\psi_\downarrow)^T$, each component 
is described by a complex bosonic field $\psi_\alpha(\textbf{r},\tau)$, 
$\psi_\sigma (\textbf{r},\tau)$.
Interaction effects will be taken into account through the intra- and 
interspecies coupling constants $g_{\alpha\alpha}$ and $g_{\alpha\sigma}$, 
respectively.
These couplings are related to the $s$-wave scattering lengths 
$a_{\alpha\alpha}$ and $a_{\alpha\sigma}$, which depend on the dimensionality of 
the system as discussed later.
Transitions between the two states are induced by an external coherent Rabi
coupling of frequency $\omega_R$.
Due to the Rabi mixing between the states, only the total number of particles is
conserved \cite{Son}. Thus it is assumed that the two components are in a state 
of chemical equilibrium $\mu$, where the chemical potential for the two 
components is the same \cite{Equal-chem-pot, Rabi1}.

In order to calculate the ground state of the mixture we obtain the grand
potential $\Omega = -\beta^{-1} \ln \mathcal{Z}$, where
$\mathcal{Z}=\int \mathcal{D}[\Psi,\Psi^*]\exp\big(-{S[\Psi,\Psi^*]/\hbar}\big)$.
To this end, we consider the superfluid phase, where a U(1) gauge symmetry of 
each component is spontaneously broken.
Then we can set $\psi_\alpha (\textbf{r},\tau) = \sqrt{n_\alpha} 
+ \eta_\alpha (\textbf{r},\tau)$, where $\sqrt{n_\alpha}$ corresponds to the
macroscopic quasicondensate (mean-field approximation), and
$n_{\alpha}=|\psi_{\alpha}|^2$ is the two- or one-dimensional quasicondensate 
density.
Although strictly Bose-Einstein condensation is prevented in low dimensionality, 
a finite-size system at a sufficiently low temperature allows for a 
quasicondensation \cite{quasi1,quasi2}.
The Gaussian fluctuations around $\sqrt{n_\alpha}$ are given by 
$\eta_\alpha (\textbf{r},\tau)$.
So, by expanding the action up to the second order in 
$\eta_\alpha (\textbf{r},\tau)$ and $\eta^*_\alpha (\textbf{r},\tau)$ 
\cite{Modes}, we arrive at the beyond-mean-field grand potential
$ \Omega(\mu,\sqrt{n_\uparrow},\sqrt{n_\downarrow})=
\Omega_0(\mu,\sqrt{n_\uparrow}, \sqrt{n_\downarrow})
+ \Omega_g(\mu, \sqrt{n_\uparrow},\sqrt{n_\downarrow})$ \cite{Andersen2}.
Here, $\Omega_0$ gives the mean-field contribution, while $\Omega_g$ takes into 
account the Gaussian fluctuations at zero temperature.
Since we want $\sqrt{n_\alpha}$ to describe the quasicondensate, then in the 
action, the linear terms in the fluctuations vanish such that $\sqrt{n_\alpha}$
really minimizes the action \cite{Stoof}. 
Thus the mean-field approximation is obtained by minimizing $\Omega_0$ with 
respect to $\sqrt{n_\alpha}$.
To get solutions from this condition, we use equal interspecies coupling 
constants $g_{11}=g_{22}=g$. In this way, two possible ground states are 
obtained, a symmetric and a polarized one \cite{Rabi1}.
Hereafter, we focus on the symmetric configuration with $n_1=n_2=n/2$.
So, we find the relation between the variational parameter $n$ and the chemical 
potential $\mu$, such that $n = 2\mu_R/g_+$, where $\mu_R=\mu+\hbar\omega_R$, 
$g_+=g(1+\epsilon)$, and $\epsilon=g_{\uparrow\downarrow}/g$. In that case, the
mean-field grand potential simplifies to \cite{Rabi}
\begin{eqnarray} 
\frac{\Omega_0}{L^D}=-\frac{\mu_R^2}{g_+}.
\end{eqnarray}
The contribution of the zero-temperature Gaussian fluctuations for the symmetric
ground-state and equal interspecies strengths reads
\begin{eqnarray} 
\frac{\Omega_g}{L^D} = \frac{1}{2}\frac{S_D}{(2\pi)^D}
\int_0^\infty dk\, k^{D-1} [E_a(k,\mu) + E_b(k,\mu)],
\label{grand-pot}
\end{eqnarray}
with $S_D=2\pi^{D/2}/\Gamma(D/2)$, the Rabi-bosonic excitations
$ E_a = [\varepsilon(k)[\varepsilon(k)+2\mu_R]]^{1/2} $, and
$ E_b = \{\varepsilon(k)[\varepsilon(k)+2 \bar{A}(\mu,\omega_R,\epsilon)] 
+ \bar{B}(\mu,\omega_R,\epsilon)\}^{1/2}$, where we have the free-particle energy
$\varepsilon(k)=\hbar^2k^2/2m$, $\bar{A}=\mu_R\Delta + 2\hbar\omega_R$, $\bar{B}
=4\hbar\omega_R(\mu_R\Delta+\hbar\omega_R)$, and $\Delta
=(1-\epsilon)/(1+\epsilon)$.

\begin{figure}[t] 
\begin{center}
\includegraphics[width=7.3cm,clip]{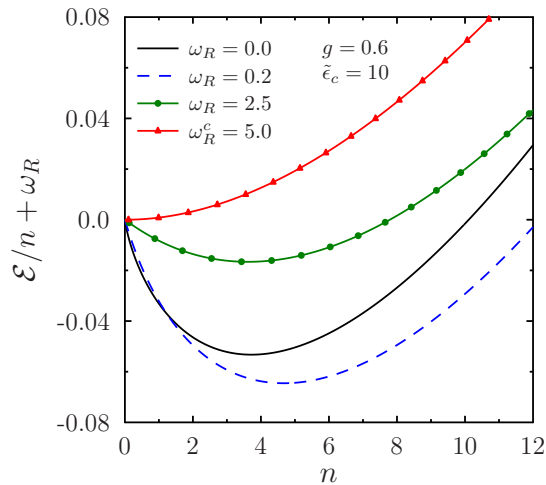}
\end{center}
\vskip -.4cm
\caption{The shifted energy per particle ${\mathcal{E}/n}+\omega_R$ vs the 
density $n$ for a two-dimensional self-bound Rabi-coupled droplet in a
Bose-Bose mixture, Eq. (\ref{droplet-2D}). 
The solid black line corresponds to a droplet phase without Rabi coupling. We 
also take into account different values of Rabi coupling, $\omega_R=0.2$ (blue, 
dashed line), $\omega_R=2.5$ (green, circles), and the critical value
$\omega^c_R=5.0$ (red, triangles).}
\label{energy-2D}
\end{figure}

\textit{Two-dimensional model.}
We first discuss the two-dimensional (2D) case with 
$g = 4\pi\hbar^2m^{-1}/\ln({4e^{-2\gamma}/a^2\kappa^2)\ll 1}$ 
\cite{scattering2D}, where $a$ is the two-dimensional scattering length,
$\gamma \approx 0.5772$ is the Euler-Mascheroni constant, and $\kappa$ is a
wave-number cutoff.
A suitable value of $\kappa$ can always be found in the weakly interacting 
regime \cite{Petrov2}. 
The repulsion (attraction) is reached for scattering lengths exponentially small
(large) compared to the mean interparticle separation.
The contribution arising from quantum fluctuations in Eq. (\ref{grand-pot}) is 
ultraviolet divergent. 
An approach to avoid this problem is through regularization methods
\cite{Hooft,dim-reg,Andersen1,quasi-condensates,Regularization}.
We obtain a fully analytical regularized momentum integration of 
Eq. (\ref{grand-pot}) (see Supplemental Material \cite{SM}).
So, in the ultradilute limit defined by $na^2\ll1$, the 2D homogeneous grand 
potential at zero temperature, including the mean-field contribution, gives
\begin{eqnarray}
\frac{\Omega}{L^2} &=& 
-\frac{\mu_R^2}{g_+}
-\frac{m}{8\pi \hbar^2}\mu_R^2
\ln{\Big(\frac{\epsilon_c}{\sqrt{e}\mu_R}\Big)} 
\nonumber \\
&-&\frac{m}{8\pi \hbar^2}\mu_R^2\Delta^2
\ln{\Big(\frac{\epsilon_c e^{-\bar{\delta}/2}}
{\mu_R \Delta + 2\hbar\omega_R}\Big)},
\label{grand-2d}
\end{eqnarray}
where $\bar{\delta}=(1-\sqrt{\bar{B}}/\bar{A})/(1+\sqrt{\bar{B}}/\bar{A})$, and 
the low-energy cutoff $\epsilon_c=\hbar^2\kappa^2/m$.
From Eq. (\ref{grand-2d}), the density $n$ is given by
\begin{eqnarray} 
n &=& \frac{2\mu_R}{g_+} + \frac{m\mu_R}{4\pi\hbar^2}
\ln \Big(\frac{\epsilon_c}{e\mu_R}\Big) +\frac{m\mu_R\Delta^2}{4\pi\hbar^2}
\Big\{\ln \Big(\frac{\epsilon_c e^{-\bar{\delta}/2}}{\bar{A}}\Big)
\nonumber \\
&-& \frac{\mu_R\Delta}{2\bar{A}} \Big[1 + \frac{2\Delta\bar{A}\hbar\omega_R\mu_R}
{\sqrt{\bar{B}}(\bar{A}+\sqrt{\bar{B}})^2}\Big]\Big\}.
\end{eqnarray}
The homogeneous ground-state energy density $\mathcal{E} =E/L^2$ also can be
read as
\begin{eqnarray} 
\mathcal{E} &=& \frac{1}{4}g_+n^2 - \hbar\omega_R n
+\frac{m}{32\pi\hbar^2}g_+^2n^2
\ln{\Big(\frac{\sqrt{e}g_+n}{2\epsilon_c}\Big)}
\nonumber \\
&+&\frac{m}{32\pi\hbar^2}g_-^2n^2
\ln{\Big(\frac{g_-n+4\hbar\omega_R}{2\epsilon_c e^{-\delta/2}}\Big)},
\label{Energy-2D}
\end{eqnarray}
with $g_-= g(1-\epsilon)$, $\delta=(1-\sqrt{B}/A)/(1+\sqrt{B}/A)$, 
$2A=g_-n + 4\hbar\omega_R$, and $B=2\hbar\omega_R(g_-n+2\hbar\omega_R)$.
Now, for simplicity, we set $m=\hbar=1$, and we consider the interesting case of 
weakly attractive inter- and weakly repulsive intraspecies interactions, where
$1/a_{\uparrow\downarrow}\ll\sqrt{n}\ll1/a$.
Notice that for $\omega_R=0$ we recover the results employed in the study of 
a two-dimensional and dilute liquidlike droplet phase.
We find that such a self-bound structure is present even in the presence of the
Rabi frequency, however, it can be unstable.
Following the lines of Ref. \cite{Petrov2}, we introduce a new energy cutoff 
$\tilde{\epsilon}_c$ including the set of coupling constants defined as
$\tilde{g}=4\pi/\ln(4e^{-2\gamma}/a^2\tilde{\epsilon}_c)$, and
$\tilde{g}_{\uparrow\downarrow}=4\pi/
\ln(4e^{-2\gamma}/a_{\uparrow\downarrow}^2\tilde{\epsilon}_c)$.
We choose $\tilde{\epsilon}_c = 4e^{-2\gamma}/a_{\uparrow\downarrow}a$ such that
$\tilde{g}^2_{\uparrow\downarrow}=\tilde{g}^2$. 
Provided that $\tilde{\epsilon}_c/\epsilon$ is not exponentially large, $g$ and
$\tilde{g}$ are equivalent. Thus from Eq. (\ref{Energy-2D}) the energy per 
particle becomes
\begin{eqnarray}
\frac{\mathcal{E}}{n}= - \omega_R
+ \frac{g^2n}{8\pi}\ln{\Big(\frac{gn+2\omega_R}
{\tilde{\epsilon}_c e^{-\tilde{\delta}/2}}\Big)},
\label{droplet-2D}
\end{eqnarray}
where 
$\tilde{\delta}= (1-\sqrt{\tilde{B}}/\tilde{A})/(1+\sqrt{\tilde{B}}/\tilde{A})$,
$\tilde{A}=gn+2\omega_R$, and $\tilde{B}=4\omega_R(gn+\omega_R)$.
The mean-field contribution of Rabi coupling gives rise to an energy per particle 
shift \cite{E1,E2}. 
However, we show that for fixed values of energy cutoff $\tilde{\epsilon}_c=10$,
and the coupling constant $g=0.6$, the increase of $\omega_R$ in the quantum
correction induces a self-bound droplet instability as plotted in Fig.
\ref{energy-2D}.
This effect is completely missed by the mean-field approximation.
We also find that the energy-cutoff dependence of Eq. (\ref{droplet-2D}) leads us
to establish that the droplet becomes unstable for 
$\tilde{\epsilon}_c \leq 2\omega_R$.

The above results apply to a purely 2D system. However, let us now comment 
briefly on the applicability of these in quasi-2D mixtures. We analyze the case 
of a harmonic confinement. In a quasi-2D system we have 
$a=2l_0\sqrt{\pi/B}e^{-\sqrt{\pi/2}l_0/a^{(3D)}-\gamma}$
\cite{scattering2D}, with $a^{(3D)}$ the three-dimensional scattering length,
$l_0$ the oscillator length of the trap in the confinement direction, and
$B \approx 0.9$.
So, the weakly interacting quasi-2D regime is reached as long as
$0<-a_{\uparrow\downarrow}^{(3D)}<a^{(3D)}\ll l_0$ \cite{Petrov2}.

\begin{figure}[t]
\begin{center}
\includegraphics[width=7.55cm,clip]{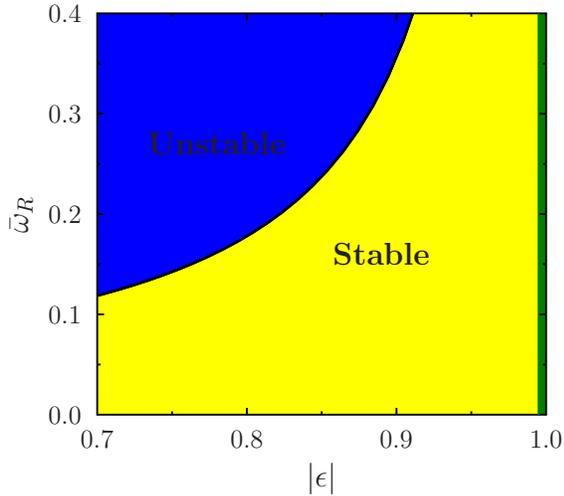}
\end{center}
\vskip -.4cm
\caption{Stability diagram of one-dimensional self-bound Rabi-coupled droplets in
Bose-Bose mixtures.
The green band shows the asymptotic behavior $|\epsilon| \rightarrow 1$.}
\label{phase-diagram}
\end{figure}

\begin{figure}[t] 
\begin{center}
\includegraphics[width=7.3cm,clip]{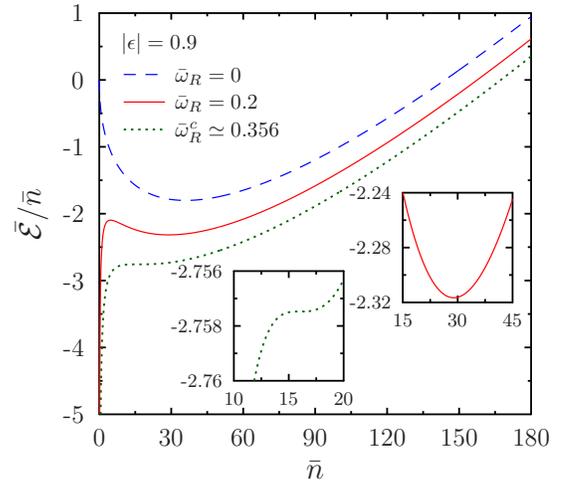}
\end{center}
\vskip -.4cm
\caption{The energy per particle $\bar{\mathcal{E}}/\bar{n}$ vs the density 
$\bar{n}$ for an one-dimensional Rabi-coupled Bose-Bose mixture, Eq. (\ref{E1D}).
The dashed line (blue) represents the droplet phase in the absence of Rabi 
coupling.
The solid line (red) considers $\bar{\omega}_R=0.2$.
The dotted line (green) corresponds to the critical value 
$\bar{\omega}_R^c\simeq 0.356$ at which the system becomes unstable. 
Inset: We present the local minimum localization of $\bar{\mathcal{E}}/\bar{n}$
and predicted by Eq. (\ref{density}).}
\label{energy-1D}
\end{figure}

\textit{One-dimensional model.}
We now turn to the weakly interacting one-dimensional scenario, which requires 
$|g|/n \ll 1$.
In particular, the 1D interparticle interaction can be well approximated by an
effective coupling constant $g=-2\hbar^2/ma$ \cite{scattering1D-1}, with $a$ the
1D scattering length.
In an attempt to get an analytical solution of Eq. (\ref{grand-pot}), and thus
insights of the underlying physics, we consider the limit of small Rabi coupling.
Then, by expanding up to the linear term in the Rabi frequency, we find that 
Eq. (\ref{grand-pot}) presents ultraviolet divergences, and an appropriate 
modification is necessary to cure these
\cite{Hooft,dim-reg,Andersen1,quasi-condensates,Regularization}. 
We employ dimensional regularization (see Supplemental Material \cite{SM}).
In this way, the regularized and homogeneous grand potential at zero temperature,
including the mean-field contribution, is
\begin{eqnarray}
\frac{\Omega}{L} &=& -\frac{\mu_R^2}{g_+}
-\frac{2}{3\pi}\Big(\frac{m}{\hbar^2}\Big)^{1/2}\mu_R^{3/2}
-\frac{2}{3\pi}\Big(\frac{m}{\hbar^2}\Big)^{1/2}(\mu\Delta)^{3/2}
\nonumber \\
&-&\frac{\omega_R}{\pi}(m\mu\Delta)^{1/2}
\Big[\Delta + \frac{1}{2}\ln{\Big(\frac{\epsilon_c}{\mu\Delta}\Big)}\Big],
\label{grand-pot-1D}
\end{eqnarray}
where $\epsilon_c\equiv\hbar^2\kappa^2e^4/64m$, with $\kappa$ an arbitrary 
wave-number scale or renormalization scale parameter.
The logarithmic contribution, although unusual in a 1D model, is proper for the 
regularization of the integrals provided by Eq. (\ref{grand-pot}), and its 
effect is discussed later.
The respective one-dimensional density is written as
\begin{eqnarray} 
n&=&\frac{2\mu_R}{g_+} + \frac{1}{\pi} \Big(\frac{m}{\hbar^2}\Big)^{1/2}
(\mu_R^{1/2}+\Delta^{3/2}\mu^{1/2})
\nonumber \\
&+& \frac{\omega_R}{2\pi}\sqrt{\frac{m\Delta}{\mu}}\Big[\Delta 
+ \frac{1}{2} \ln\Big({\frac{\epsilon_c}{e\mu\Delta}}\Big)\Big],
\end{eqnarray}
and the ground-state energy density is given by
\begin{eqnarray} 
\mathcal{E} &=& \frac{1}{4}g_+n^2 - \hbar \omega_R n
-\frac{1}{3\pi}\Big(\frac{m}{2\hbar^2}\Big)^{1/2}
n^{3/2}(g_+^{3/2} + g_-^{3/2})
\nonumber \\
&-& \frac{\omega_R}{\sqrt{2}\pi}(mg_-n)^{1/2} \Big[\Delta 
+ \frac{1}{2}\ln{\Big(\frac{2\epsilon_c}{g_-n}\Big)} \Big].
\label{E}
\end{eqnarray}
From a mean-field viewpoint, and in the absence of Rabi coupling, the condition
$0<\epsilon<1$ is employed in order to avoid phase separation
\cite{Pit-Strin,Peth-Smith}.
However, for $\epsilon>1$, the inclusion of Rabi coupling gives rise to an 
effective attraction between the species, which can drive the immiscible 
configuration into a miscible state \cite{Miscibility}.
Instead, from energy (\ref{E}), we find that for repulsive intraspecies and 
interspecies interactions ($g,g_{\uparrow\downarrow}>0$) with $\epsilon>1$, the 
quantum corrections destabilize the system.
On the other hand, in the regime of repulsive ($g>0$) intraspecies and attractive 
($g_{\uparrow\downarrow}<0$) interspecies interactions for $\epsilon<-1$,
the system becomes fully attractive and unstable, contrasting thus the 
three-dimensional one \cite{Rabi}. In such a 3D scenario, the resulting energy 
density displays an interplay between an attractive mean-field term 
$\propto -n^2$ and a repulsive beyond-mean-field correction $\propto n^{5/2}$. 
This feature opens the door to the possibility of observing a droplet phase.
We now focus on the regime of repulsive intraspecies and attractive interspecies 
interactions with $|\epsilon|\sim 1$ \cite{Petrov2}.
By using $g=2\hbar^2/m|a|$, 
$g_{\uparrow\downarrow}=-2\hbar^2/m |a_{\uparrow\downarrow}|$ 
\cite{scattering1D-1}, and by taking $E_B=\hbar^2/ m |a|^2$ as the energy unit, 
we define $\bar{n}=n|a|$, $\bar{\omega}_R=\hbar\omega_R/E_B$, and 
$\bar{\epsilon}_c = \epsilon_c/E_B$. 
Thus we get the scaled energy per particle of ultradilute and uniform 
Rabi-coupled mixtures 
$\bar{\mathcal{E}}/\bar{n}=\mathcal{E}/(\bar{n} E_B/|a^{1D}|)$ as
\begin{eqnarray} 
\frac{\bar{\mathcal{E}}}{\bar{n}} = \frac{1}{2}\epsilon_-\bar{n}
- \bar{\omega}_R -\frac{4\sqrt{2}}{3\pi}\bar{n}^{1/2}
- \frac{2\sqrt{2}}{\pi}\frac{\bar{\omega}_R}{\epsilon_-\bar{n}^{1/2}},
\label{E1D}
\end{eqnarray}
where $\epsilon_-=1-|\epsilon|\rightarrow 0$. In this regime, we find that 
\begin{eqnarray} 
\frac{1}{\epsilon_-} 
\gg \frac{1}{4}\ln\Big(\frac{\bar{\epsilon_c}}{4\bar{n}}\Big).
\end{eqnarray}
So, we check that the logarithmic contribution of energy (\ref{E}) does not
affect the results, and hence it is neglected \cite{cutoff}.
This expression accounts for the beyond-mean-field attractive correction to the 
ground-state energy.
In the absence of Rabi coupling such a mixture gets into a pure dilute 
liquidlike droplet state.
Here, we find that such energy per particle with $\bar{\omega}_R \neq 0$
displays that the Rabi coupling shifts the ground-state energy of the self-bound
droplet, and also it leads to the possibility of obtaining a quantum mechanical
instability of the mixture.
By minimizing Eq. (\ref{E1D}) with respect to density, we find two 
possible physical solutions which are written as
\begin{eqnarray} 
\bar{n}_j=\frac{128}{81\pi^2\epsilon_-^2}\Big[\cos
\Big[\frac{1}{3}\cos^{-1}\theta(\bar{\omega}_R,|\epsilon|)-\frac{2\pi}{3}j\Big]
+ \frac{1}{2}\Big]^2,
\label{density}
\end{eqnarray}
where $\theta = 1-729\pi^2\bar{\omega}_R \epsilon_-/ 128$, and $j=0,1$.
For $j=0$ we have the equilibrium density, hereafter called $\bar{n}_0$, while
for $j=1$ the solution is a local maximum.
From $\bar{n}_0$ we find a stable mixture for 
$0< \bar{\omega}_R(1-|\epsilon|)\leq 256/729\pi^2$, as plotted in Fig. 
\ref{phase-diagram}.
The green band shows the asymptotic behavior $|\epsilon|\rightarrow 1$.
This diagram could be understood as a result of the combined attraction-repulsion
effect of the different contributions in Eq. (\ref{E1D}).
As a particular case we display in Fig. \ref{energy-1D} the energy per particle
given by Eq. (\ref{E1D}) as a function of the density. We consider a fixed value
of $|\epsilon|=0.9$. The increase of $\bar{\omega}_R$ shows as the local minimum
disappears when this frequency exceeds the critical value of 
$\bar{\omega}_R^c \simeq 0.356$, as predicted by Eq. (\ref{density}).
Beyond this critical parameter we have an effective attraction between atoms and
the ground state becomes unstable.
This is a similar mechanism to the instability by collapse in a single-component
3D BEC \cite{Pit-Strin,Peth-Smith}.
We also calculate the spinodal region of the mixture $\bar{n}̣^\mathrm{sp}$,
which is defined by the condition 
$\partial^2\bar{\mathcal{E}}/\partial \bar{n}^2=0$. 
This spinodal density has the same form of Eq. (\ref{density}), provided that 
$\epsilon_- \rightarrow 4\epsilon_-/3$ and
$\omega_R \rightarrow 4\omega_R/9$. So, the mixture is thus metastable for 
$\bar{n}^{\mathrm{sp}}<\bar{n}<\bar{n}_0$.
In the case of $|\epsilon|=0.9$ and $\omega_R=0.2$, we find a metastable mixture
for $18.085<\bar{n}<28.945$.
From the ground-state energy density of the Rabi-coupled droplet, 
Eq. (\ref{E1D}), we also get some relevant quantities as the equilibrium
chemical potential $\bar{\mu}_0 = \bar{\mu}(\bar{n}=\bar{n}_0)$,
\begin{eqnarray} 
\bar{\mu}_0 = \epsilon_-\bar{n}_0 - \bar{\omega}_R 
-\frac{2\sqrt{2}}{\pi}\bar{n}_0^{1/2}
-\frac{\sqrt{2}}{\pi}\frac{\bar{\omega}_R}{\epsilon_-\bar{n}_0^{1/2}},
\end{eqnarray}
and the corresponding scaled speed of sound $\bar{c}_0\equiv c_0/\sqrt{E_B/m}$,
given by
\begin{eqnarray} 
\bar{c}_0=\Big(\epsilon_-\bar{n}_0- \frac{\sqrt{2}}{\pi}\bar{n}_0^{1/2}
+ \frac{1}{\sqrt{2}\pi}
\frac{\bar{\omega}_R}{\epsilon_-\bar{n}_0^{1/2}}\Big)^{1/2}.
\end{eqnarray}

Let us now comment on the applicability of these results in quasi-1D mixtures.
In a quasi-1D system where a harmonic radial confinement with oscillator length 
$l_\perp$ is assumed , the relation between $a^{(3D)}$ and the one-dimensional 
scattering length is $a = -l_\perp(l_\perp-Ca^{(3D)})/a^{(3D)}$
\cite{scattering1D-1}, with  $C \simeq 1.0326$.
So, an effective weakly interacting quasi-1D regime is obtained as long as the 
radial confinement, fixed by the radial oscillator length, is much greater than
the three-dimensional scattering lengths, i.e.,
$l_{\perp} \gg a^{(3D)} > -a_{\uparrow\downarrow}^{(3D)}>0$.

\textit{Conclusion.}
We study ultracold and ultradilute weakly interacting low-dimensional 
Rabi-coupled Bose-Bose mixtures beyond the mean-field approximation.
In particular, we focus on the case where the interspecies interactions are 
weakly attractive and the intraspecies ones are weakly repulsive.
Our results show that in these regimes such mixtures manifest the formation of a 
liquidlike Rabi-coupled droplet phase including a quantum instability
encoded in the Rabi frequency.
Remarkably, in the 1D mixture we provide an analytical equilibrium density, and
also identify the spinodal region of the self-bound droplets.
Furthermore, in the stable liquidlike state, we also calculate the chemical 
potential, and the speed of sound.
The above general features of Rabi-coupled droplets are directly linked to the 
crucial nature of the quantum fluctuations, and hence may stimulate and play an 
interesting role in future experiments.
It also may be interesting to verify the validity of our results by means of the
recent developments on quantum Monte Carlo methods 
\cite{Montecarlo-Bose-Bose1,Montecarlo-Bose-Bose2}.

\clearpage\section{Supplemental Material}
\vskip -.25cm
\textit{Grand potential in two dimensions.} From Eq. (\ref{grand-pot}) in main
text, the grand potential in two dimensions takes the form
\begin{eqnarray} 
\frac{\Omega_g}{L^2}= I_a + I_b,
\end{eqnarray}
where, 
\begin{eqnarray} 
I_a = \int_0^\infty \frac{dk}{4\pi} k
\sqrt{\frac{\hbar^2k^2}{2m}\Big(\frac{\hbar^2k^2}{2m} +2\mu_R\Big)},
\end{eqnarray}
and
\begin{eqnarray} 
I_b = \int_0^\infty \frac{dk}{4\pi} k
\sqrt{\frac{\hbar^2k^2}{2m}\Big(\frac{\hbar^2k^2}{2m} +2\bar{A}\Big)+\bar{B}}.
\end{eqnarray}
To calculate $I_a$, we employ dimensional regularization in the modified minimal
subtraction scheme $\mathrm{\overline{MS}}$-scheme. In a similar way, this 
integral also is calculated by means of convergence-factor regularization (CFR) 
\cite{Regularization-SM}, with same results.
To deal with the second term we use the CFR method. 
Thus, $I_a$ is extended to a noninteger and generic $D=2-2\bar{\varepsilon}$
dimension, and the limit $\bar{\varepsilon}\rightarrow 0$ is applied at the end 
of the calculation. So
\begin{eqnarray} 
I_a &=& -\frac{m\mu_R^2}{4\pi^{3/2}\hbar^2}
\Big(\frac{\hbar^2\pi}{m\mu_R}\Big)^{\bar{\varepsilon}}
\Big(\frac{e^{\gamma}\kappa^2}{4\pi}\Big)^{\bar{\varepsilon}}
\Gamma\Big(\frac{3}{2}\Big)
\nonumber \\
&\times&\Big[\frac{1}{\bar{\varepsilon}}
+\ln{\Big(\frac{4}{e^{\gamma + 1/2}}}\Big)\Big],
\end{eqnarray}
where $\kappa$ is an arbitrary wavenumber scale or renormalization scale
parameter.
The factor $(e^{\gamma}/4\pi)^{\bar{\varepsilon}}$ is introduced so that, after 
minimal subtraction of the poles in $\bar{\varepsilon}$, $\kappa$ coincides with
the renormalization scale of the $\mathrm{\overline{MS}}$-scheme
\cite{dim-reg-SM}.
Then by expanding $I_a$ in terms of the parameter $\bar{\varepsilon}$, we get
\begin{eqnarray} 
I_a = -\frac{m\mu_R^2}{8\pi\hbar^2}\Big[\frac{1}{\bar{\varepsilon}}
+\ln\Big({\frac{\hbar^2\kappa^2}{m\mu_R\sqrt{e}}}\Big)
+ \mathcal{O}(\bar{\varepsilon})\Big].
\end{eqnarray}
By performing the appropriate counterterms subtraction to remove the pole in
$\bar{\varepsilon}$
\cite{counterterms1,counterterms2,counterterms3,Andersen1-SM,Andersen3-SM}, 
and by using the limit $\bar{\varepsilon}\rightarrow 0$, we obtain the second
term in Eq. (\ref{grand-2d}) of main text.

Regarding to $I_b$, we have
\begin{eqnarray} 
I_b &=& \frac{m}{2\pi\hbar^2}\int dz \Big[z\sqrt{z^4+2\bar{A}z^2+\bar{B}}
\nonumber \\
&-&\Big(z^3+\bar{A}z+\frac{\bar{B}-\bar{A}^2}{2z}\Big)\Big],
\end{eqnarray}
with $z^2=\hbar^2k^2/2m$.
The counterterms are determined by expanding the Rabi excitation $E_b$ at high
momentum \cite{Rabi-SM}.
Solution of this integral is given by the last term in Eq. (\ref{grand-2d}) of
main text. 

\textit{Grand potential in one dimension.}
In the one-dimensional case and employing the approximation of small 
Rabi-coupling, Eq. (\ref{grand-pot}) of main text is written as
\begin{eqnarray} 
\frac{\Omega_g}{L}&=& I_a + I_1 + I_2 + I_3,\\ \nonumber
\end{eqnarray}
where the integrals $I_a$, $I_1$, and $I_2$ are calculated by means of 
dimensional regularization (with same results using CFR). Thus
\begin{eqnarray} 
I_a &=& \int_0^\infty \frac{dk}{2\pi}
\sqrt{\frac{\hbar^2k^2}{2m}\Big(\frac{\hbar^2k^2}{2m} +2\mu_R\Big)}
\nonumber \\
&=& -\frac{2}{3\pi}\Big(\frac{m}{\hbar^2}\Big)^{1/2}\mu_R^{3/2},
\end{eqnarray}
\begin{eqnarray} 
I_1 &=& \int_0^\infty \frac{dk}{2\pi}
\sqrt{\frac{\hbar^2k^2}{2m}\Big(\frac{\hbar^2k^2}{2m} +2\mu\Delta\Big)}
\nonumber \\
&=& -\frac{2}{3\pi}\Big(\frac{m}{\hbar^2}\Big)^{1/2}(\mu\Delta)^{3/2},
\end{eqnarray}
and,
\begin{eqnarray} 
I_2 &=& \frac{1}{\pi}\omega_R(\Delta + 2)(m\mu\Delta)^{1/2}\int_0^\infty 
\frac{x}{\sqrt{x^2+1}}
\nonumber \\
&=& -\frac{\omega_R}{\pi}(\Delta + 2)(m\mu\Delta)^{1/2}.
\end{eqnarray}
with $x^2=\hbar^2k^2/4m\mu\Delta$.
Integration over momentum of $I_3$ it is still divergent, so we also use the 
$\mathrm{\overline{MS}}$-scheme.
Then through $D=1-2\bar{\varepsilon}$, we get
\begin{eqnarray} 
I_3 &=&\frac{1}{\pi}\omega_R(m\mu\Delta)^{1/2}\int_0^\infty 
\frac{1}{x\sqrt{x^2+1}} 
\nonumber\\
&=& \frac{\omega_R}{2\pi}(m\mu\Delta)^{1/2}
\Big(\frac{\pi\hbar^2}{m\mu\Delta}\Big)^{\bar{\varepsilon}}
\Big(\frac{e^{\gamma}\kappa^2}{4\pi}\Big)^{\bar{\varepsilon}}
\frac{\Gamma(-\bar{\varepsilon})\Gamma(\frac{1}{2}+\bar{\varepsilon})}
{\Gamma(\frac{1}{2}-\bar{\varepsilon})}
\nonumber\\
&=& \frac{\omega_R}{2\pi}(m\mu\Delta)^{1/2}\Big[-\frac{1}{\bar{\varepsilon}} 
+ \ln\Big(\frac{64m\mu\Delta}{\hbar^2\kappa^2}\Big)\Big]
+ \mathcal{O}(\bar{\varepsilon})
\end{eqnarray}
By taking into account the appropriate counterterms subtraction to remove the 
pole in $\bar{\varepsilon}$, and applying the limit
$\bar{\varepsilon}\rightarrow 0$, we obtain the last term of 
Eq. (\ref{grand-pot-1D}) in main text.
$I_3$ also is solved by means of CFR and we find equivalence between both methods
as long as $\kappa_{\mathrm{\overline{MS}}}^2= 4q_\mathrm{{CFR}}^2$, with
$q_{\mathrm{CFR}}$ a low-wavenumber cutoff.


\begin{thebibliography}{99}
\bibitem{cutoff} Regarding the energy cutoff in the logarithmic term, we also
consider the scaled energy per particle in a general context. We find that for 
$0<\epsilon\ll1$, the mean-field contribution dominates over the quantum
corrections. So, the effect of $\bar{\epsilon}_c$ is not relevant either.
We confirm our results by using 
$3\cdot10^{-1} \leq \bar{\epsilon}_c \leq 5\cdot10^{3}$.
\end{thebibliography}
\end{document}